\documentclass{jpsj-suppl}
\usepackage{txfonts} %Please comment out this line unless the txfonts package is availabe in your LaTeX system.
\usepackage{dcolumn}% Align table columns on decimal point
\usepackage{bm}% bold math

\usepackage{graphicx}% Include figure files
\usepackage{amsmath}
\usepackage{amssymb}
\usepackage{color}

\def\beq{\begin{equation}}
\def\eeq{\end{equation}}
\def\be{\begin{eqnarray}}
\def\ee{\end{eqnarray}}

\title{Green's function Monte Carlo calculations of the electromagnetic and neutral-weak response
functions in the quasi-elastic sector}

\author{A. \textsc{Lovato}$^{1}$ }

\inst{$^{1}$Physics Division, Argonne National Laboratory, Argonne, IL 60439 \\
}
\email{lovato@anl.gov}

\recdate{January 31, 2016}

\abst{A quantitative understanding of neutrino-nucleus interactions is demanded
to achieve precise measurement of neutrino oscillations, and hence the determination
of their masses. In addition, next generation detectors will be able to detect
supernovae neutrinos, which are likely to shed some light on the open questions on the
dynamics of core collapse.
In this context, it is crucial to account for two-body meson-exchange currents along within
realistic models of nuclear dynamics. We summarize our progresses towards the construction
of a consistent framework, based on the Green's function Monte Carlo method, that can be 
exploited  to accurately describe neutrino interactions with atomic nuclei in the quasi-elastic
sector.}

\kword{Lepton-nucleus scattering, Meson exchange currents, Quantum Monte Carlo}

\begin{document}
\maketitle

\section{Introduction}
\label{intro}
The description of neutrino interactions with nuclei, beside being fundamental for understanding the coupling of electroweak interactions to nuclei, provides an essential input for neutrino-oscillation experiments, such as DUNE \cite{LBNE:2013}, which plan to measure neutrino oscillation parameters, the neutrino mass hierarchy, and to determine the charge-conjugation and parity (CP) violating phase. Such experiments make use of nuclear targets as detectors that, while allowing for a substantial increase in the event rate, demand accurate estimates of the neutrino-nucleus cross section along with the associated theoretical error. The latter is critical, as the uncertainties in neutrino-nucleus cross section propagate into the systematic error of the neutrino oscillation parameters. It has to be noted that a similar issue also occurs in neutrino-less double beta decay experiments, the systematic uncertainty of which is dominated by the nuclear matrix element.

The MiniBooNE collaboration has reported a measurement of the charged-current quasielastic (CCQE) neutrino-carbon inclusive double differential cross section, exhibiting a large excess with respect to the predictions of relativistic Fermi gas model \cite{Aguilar:2008}. It has been suggested that this discrepancy might be ascribed to the occurrence of events with two particle-two hole final states \cite{Martini:2009,Nieves:2011} not taken into account by the relativistic Fermi gas model. Within a realistic model of nuclear dynamics, their occurrence arises naturally owing to two-body meson-exchange currents (MEC) and/or correlations induced by nuclear interactions.

Since their first application more than fifty years ago to a classical system of hard disks, Monte Carlo methods have been found indispensable for many different simulations of quantum many-body systems. The main advantage of quantum Monte Carlo is that it is {\it statically exact}, as it allows for solving the time-independent
Schr\"odinger equation of many-body Hamiltonians providing accurate estimates of the gaussian error of the calculation. For light nuclei (s- and p-shell nuclei up to $^{12}$C), quantum Monte Carlo and, in particular, Green's Function Monte Carlo (GFMC) methods have been exploited to carry out first-principle calculations of nuclear properties, based on realistic, albeit phenomenological, Hamiltonians including two- and three-nucleon potentials, and consistent one- and two-body MEC currents\cite{Carlson:2015}. Within this phenomenological model of nuclear dynamics, GFMC has been also successfully applied to study of the electroweak response of light nuclei \cite{Lovato:2013,Lovato:2013b,Lovato:2015}.

\begin{figure}
% Use the relevant command for your figure-insertion program
% to insert the figure file.
\centering
\includegraphics[width=8.5cm,clip]{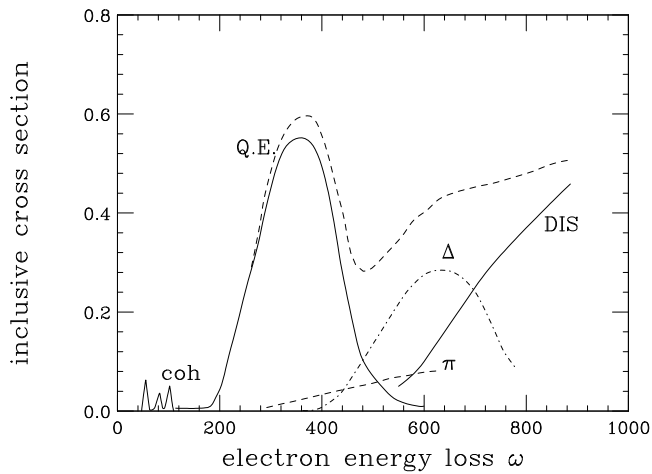}
\caption{Schematic representation of the inclusive electron-nucleus cross section at beam energy around 1 GeV, as a function
of energy loss, from Ref. \cite{Benhar:2006wy}.}
\label{fig-1}       % Give a unique label
\end{figure}

Because neutrino beams are always produced as secondary decay products, they are not monochromatic, as their energy is broadly distributed. As a consequence, the observed cross section for a given energy and angle of the outgoing lepton includes contributions from energy- and momentum-transfer regions where different mechanisms are at play. Such mechanisms can be best identified in electron-scattering experiments, in which the energy of the incoming electron is precisely known. The typical behavior of the double differential inclusive cross section of process
\begin{equation}
e+A\to e^\prime + X\, ,
\end{equation}
for a beam energy around 1 GeV where only the outgoing electron is detected is shown in Figure~\ref{fig-1}, taken from Ref. \cite{Benhar:2006wy}. The target nucleus in its ground state and the undetected hadronic final state are denoted by $A$ and $X$, respectively.
At small energy transfer the structure of the low-lying energy spectrum and collective effects are important. In the quasielastic peak region, the cross section is dominated by scattering off individual nucleons although a significant contribution from nucleon pairs is also present. The $\Delta$-resonance region is characterized by one or more pions in the final state, while at large value of the energy transfer the deep inelastic scattering contribution is the largest.

The initial state of the nucleus does not depend upon the momentum transfer of the process. Hence, it can be accurately described within nonrelativistic many-body theory (NMBT). On the other hand, the final state does depend on the momentum transfer and the nonrelativistic approximation can be safely applied only in the region of low and moderate momentum transfer, corresponding to $|\mathbf{q}|\lesssim 500$ MeV.  The treatment of the region of high momentum transfer requires a theoretical approach in which the accurate description of the nuclear ground state provided by NMBT is combined with a relativistically consistent description of both the final state and the nuclear current. In this proceedings we will focus on the low and moderate momentum transfer regions. The framework in which two-body currents and nuclear correlations induced by realistic hamiltonians are consistently accounted for in a fully relativistic fashion is extensively discussed in Ref. \cite{Rocco:2016}.

\section{Lepton-Nucleus cross section}
The double differential inclusive cross-section of the electromagnetic (EM) and neutral-current (NC) lepton-nucleus transition $\ell+A\to\ell^\prime +X$ is proportional to the product of the leptonic and the hadronic tensors,
\begin{equation}
\frac{d\sigma}{d\Omega_\ell d E_\ell} \propto L_{\mu\nu}W^{\mu\nu}\ .
\end{equation}
The leptonic tensor $L_{\mu\nu}$ is completely determined by lepton kinematics, whereas the hadronic tensor, 
\begin{align}
\label{hadronictensor}
W_{\mu\nu}&= \sum_X \,\langle 0 | {J_\mu}^\dagger | X \rangle \,
      \langle X | J_\nu | 0 \rangle \;\delta^{(4)}(p_0 + q - p_X) \ ,
\end{align}
containing all the information on strong interaction dynamics, describes the nuclear response of the target to the transition currents ${J_\mu}$ from the initial state $|0\rangle$, with four momenta $p_0$, to the final state $|X\rangle$, with four momenta $p_X$. 

The nuclear current consists of one- and two-nucleon contributions, the latter arising from processes in which the interaction with the beam particle involves a meson exchanged between the target nucleons are usually referred to as meson-exchange currents (MEC) 
\begin{equation}
\label{def:curr}
J^\mu = J_1^{\,\mu} + J_2^{\,\mu} = \sum_i j^{\,\mu}_i + \sum_{\rm j>i} j^{\,\mu}_{ij} \ .
\end{equation}

In the moderate momentum transfer regime, both the initial and the final states appearing in  Eq. (\ref{hadronictensor}) 
are eigenstates of the nonrelativistic nuclear hamiltonian $H$
\begin{equation}
\label{schroedinger}
H |0\rangle=E_0|0\rangle \ \ \ \ , \ \ \ \  H|X\rangle=E_X|X\rangle \ .
\end{equation}
The nuclear Hamiltonian describing the dynamics of the point like nonrelativistic nucleons is given by
\begin{align}
H = \sum_{i=1}^{A} \frac{{\bf p}_i^2}{2m} + \sum_{j>i=1}^{A} v_{ij}
 + \sum_{k>j>i=1}^A V_{ijk} \ .
\label{eq:nucl_h}
\end{align}
In the above equation, ${\bf p}_i$ is the momentum of the $i$-th nucleon, while $v_{ij}$ and $V_{ijk}$ are the two- and three-nucleon potentials, respectively.

For nuclei as large as $^{12}$C, the ground state wave function can be obtained via the GFMC approach \cite{kalos:1962,grimm:1971}. It has to be remarked that, within the limits of applicability of NMBT, GFMC is truly an ab initio approach, allowing to perform {\it statistically exact} calculations of a number of nuclear properties.

A key feature of the description of neutrino-nucleus interactions at low and moderate momentum transfer is the possibility of employing a set of electroweak charge and current operators consistent with the Hamiltonian of Eq.(\ref{eq:nucl_h}). The nuclear electromagnetic current,   $J^\mu_{\rm em} \equiv (J^0_{\rm em},{\bf J}_{\rm em})$, trivially related to the vector component of the weak current, is constrained by $H$ through the continuity equation \cite{Riska89}
\begin{equation}
\label{continuity}
{\boldsymbol \nabla} \cdot {\bf J}_{\rm em} + i [H,J^0_{\rm em}] = 0 \ .
\end{equation}
Since the NN potential $v_{ij}$ does not commute with the charge operator $J^0_{\rm em}$, the above equation implies that $J^\mu_{\rm em}$ involves two-nucleon contributions, as shown in Eq.\eqref{def:curr}.

The one-body electroweak operator is obtained from a non relativistic expansion of the covariant single-nucleon currents, while the two-body charge and current operators that we employed in our calculations are derived within the conventional meson-exchange formalism \cite{Marcucci:2000, Marcucci:2005}. Nonrelativistic MEC have been used in analyses of a variety of electromagnetic moments and  electroweak transitions of s- and p-shell nuclei at low and intermediate values of energy and momentum transfers \cite{Pervin:2007,Marcucci:2008,Schiavilla:2002,Marcucci:2011,Wiringa:1998,Marcucci:1998,Viviani:2007}, improving on the description of the experimental data with respect to the one-body approximation. 

\section{Quantum Monte Carlo calculations of the Euclidean response functions}
\label{sec-3}
In the low and moderate momentum-transfer regimes, the nuclear cross section can be rewritten in terms of the response functions $R_{\mu\nu}(q,\omega)$, obtained from Eq.\eqref{hadronictensor} replacing the components of the current operator with their expressions obtained in the nonrelativistic limit
\begin{equation}
R_{\alpha\beta}(q,\omega)= \sum_X \,\langle 0 | {J_\alpha}^\dagger(\mathbf{q},\omega) | X \rangle \,
      \langle X | J_\beta(\mathbf{q},\omega) | 0 \rangle \;\delta(E_0 + \omega - E_X) \ .
\end{equation}

Even in the quasi-elastic region and for moderate values of the momentum transfer, where the consequences of the nucleon's substructure on nuclear dynamics can be subsumed into effective many-body potentials and currents, the calculation of the response functions remains extremely difficult, as it requires knowledge of the whole excitation spectrum of the nucleus and inclusion in the electroweak currents of one- and many-body terms. Integral properties of the response functions can be studied by means of sum rules, which are obtained from ground-state expectation values of appropriate combinations of the current operators, thus avoiding the need for computing the nuclear excitation spectrum. Ab initio GFMC calculations of the electroweak sum rules in $^{12}$C have been recently reported in Refs. \cite{Lovato:2013,Lovato:2013b}. These calculations have demonstrated that a large fraction ($\simeq 30\%$) of the strength in the transverse response arises from processes involving two-body currents, and that interference effects between the matrix elements of one- and two-body currents play a major role \cite{Benhar:2015ula}. These effects are typically only partially, or approximately, accounted for in existing perturbative or mean-field studies \cite{Martini:2009,
Martini:2010,Nieves:2011,Amaro:2011}.

We have used the GFMC method to calculate the imaginary-time response function--the so-called Euclidean response function--to the electromagnetic and neutral weak currents of $^{4}$He and $^{12}$C at low and moderate values of the momentum transfer. The Euclidean response function is defined as the Laplace transform
of the response
\begin{equation}
E_{\alpha\beta}(q,\tau) = C_{\alpha\beta}(q)\int_{\omega_{\rm th}}^\infty d\omega\,
 e^{-\tau \omega} R_{\alpha\beta}(q,\omega) \ ,
\label{eq:laplace_def}
\end{equation}
where $\omega_{\rm th}$ is the inelastic threshold and the $C_{\alpha\beta}$ are $q$-dependent normalization factors. In $R_{\alpha\beta}(q,\omega)$ the
$\omega$-dependence enters via the energy-conserving $\delta$-function and the dependence on the four-momentum transfer $Q^2=q^2-\omega^2$ of the
electroweak form factors of the nucleon and of the $N$-to-$\Delta$ transition in the currents. Once the latter dependance has been removed, as described in Ref. \cite{Lovato:2015}, the Euclidean response can be expressed as a ground-state expectation value
\begin{equation}
\frac{E_{\alpha\beta}(q,\tau)}{C_{\alpha\beta}(q)}= \frac{\langle 0| J^\dagger_{\alpha}({\bf q}) e^{-(H-E_0)\tau} 
J_{\beta}({\bf q}) |0\rangle}{\langle 0| e^{-(H-E_0)\tau}|0\rangle}\, .
\label{eq:euc_me}
\end{equation}
In the above equation $H$ is the nuclear Hamiltonian of Eq. (\ref{eq:nucl_h}) (here, the AV18+IL7 model is adopted), $\tau$ is the imaginary-time, and $E_0$ is a trial energy to control the normalization. The calculation of the above matrix element is carried out with GFMC methods~\cite{Carlson:1992} similar to those used in projecting out the exact ground state of the Hamiltonian from a trial wave function. 

\begin{figure}
% Use the relevant command for your figure-insertion program
% to insert the figure file.
\centering
\includegraphics[width=8.5cm,clip]{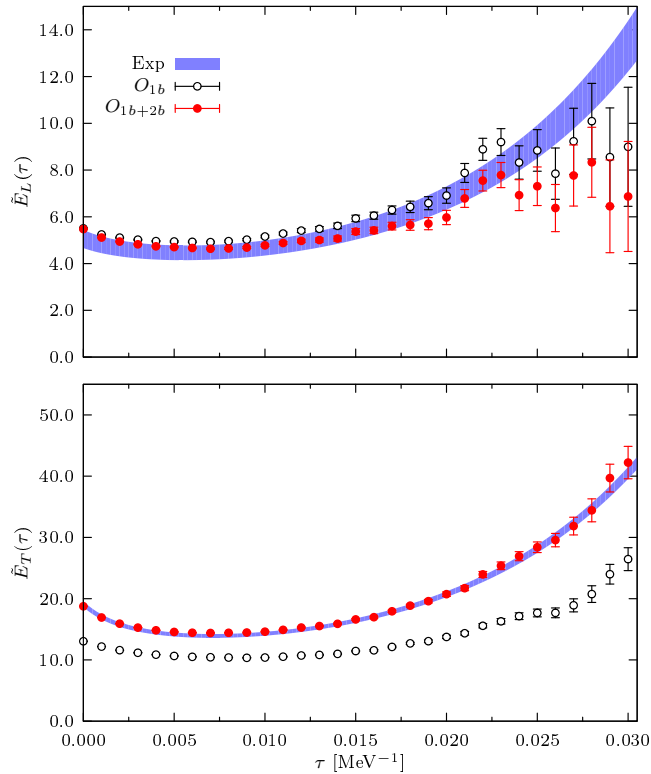}
\caption{Euclidean electromagnetic longitudinal (upper panel) and transverse (lower panel) response functions of $^{12}$C at $q = 570$ MeV (from Ref. \cite{Lovato:2015}). The theoretical predictions obtained including only one-body and both one- and two-body transition operators are represented by open circles and solid circles, respectively. The shaded band refers to the Euclidean response extracted from Ref. \cite{Jourdan:1996}.}
\label{fig:12C_euc}      % Give a unique label
\end{figure}

It has to be noted that evaluating the matrix element of Eq. (\ref{eq:euc_me}) for a nucleus as large as $^{12}$C is computationally nontrivial, as it requires computer capabilities and resources available at the present time to be stretched to their very limit.

In Figure \ref{fig:12C_euc}, taken from Ref. \cite{Lovato:2015}, the electromagnetic longitudinal and transverse Euclidean response function of $^{12}$C are compared to the ones obtained from the analysis of the world data carried out by Jourdan \cite{Jourdan:1996} represented by the shaded band. In order to better show the large $\tau$ behavior, the scaled response functions $\tilde E_{\alpha\beta}(\tau)\equiv \exp[\tau q^2/(2m)] E_{\alpha\beta}$, that would be a constant for an isolated proton, are displayed . 

In the longitudinal case, destructive interference between the matrix elements of the one- and two-body charge operators reduces, albeit slightly, the one-body response. On the other hand, two-body MEC contributions substantially increase the one-body electromagnetic transverse response function. This enhancement is effective over the whole imaginary-time region we have considered, with the implication that excess transverse strength is generated by two-body currents not only at energies larger than the one corresponding to the quasi-elastic peak, but also in the quasi-elastic and threshold regions. The full predictions for the transverse Euclidean response functions is in excellent agreement with the experimental data.

\section{Analytic continuation of the Euclidean response functions}
\label{sec-4}

The inversion of a Laplace transform subject to statistical Monte Carlo errors, needed to retrieve the energy dependence of the responses, is long known to involve severe difficulties. However, there are techniques developed in condensed matter theory and other contexts, that seem to have successfully overcome the inherent ill-posed nature of the problem. Indeed one of those, known as {\it maximum entropy technique}, has recently been used to perform stable inversions of the $^4$He electromagnetic Euclidean response \cite{Lovato:2015}.  It has to be noted that the fact that we are interested in the (smooth) response around the quasi-elastic peak rather than isolated peaks makes the inversion somewhat more practical. The maximum-entropy method is based on Bayesian statistical inference: the ``most probable'' response function is the one that maximizes the posterior probability $Pr[R|E]$, i.e., the conditional probability of obtaining the response function $R(\omega)$ given the Euclidean response function $E(\tau)$.  Maximum entropy technique is based on the fact that, since the response function is positive definite and normalizable, it can be interpreted as yet another probability function. The principle of maximum entropy states that the values of a probability function are to be assigned by maximizing the entropy, defined as
\begin{equation}
S=\int d\omega \left[R(\omega)-M(\omega)-R(\omega)\ln[R(\omega)/M(\omega)]\right]\, .
\end{equation}
The entropy measures how much the response function differs from the the model (prior) $M(\omega)$: it vanishes when $R(\omega) = M(\omega)$, and is negative when $R(\omega)\neq M(\omega)$. In principle the inversion of the Euclidean response depends upon the choice of the model. In order to study the sensitivity of the solution on the choice of the prior, the Authors of Ref. \cite{Lovato:2015} have used two non-informative default models: the flat model and a simple gaussian centered in $\omega=0$. 

\begin{figure}
% Use the relevant command for your figure-insertion program
% to insert the figure file.
\centering
\includegraphics[width=8.5cm,clip]{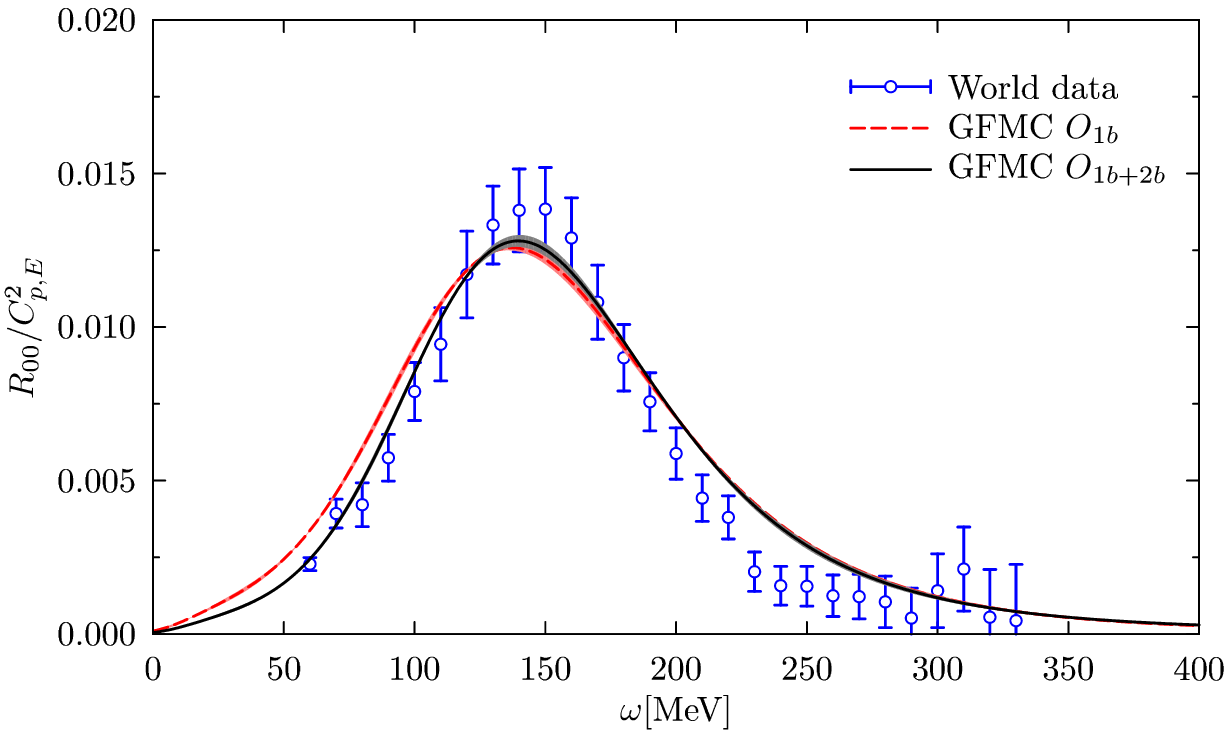}
\includegraphics[width=8.5cm,clip]{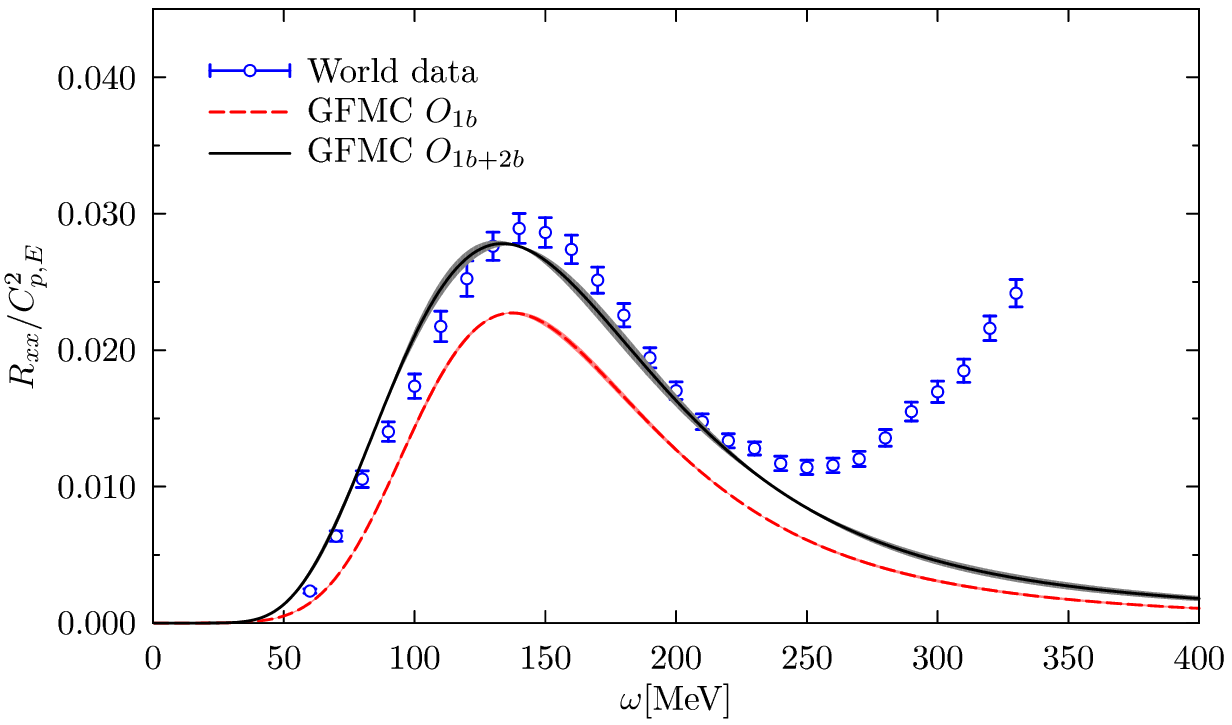}
\caption{Electromagnetic longitudinal (upper panel) and transverse (lower panel) response functions of $^4$He at $q = 500$ MeV. Experimental data are from Ref.\cite{Carlson:2002}.}
\label{fig:4he_xx}     
\end{figure}

The $^4$He electromagnetic longitudinal and transverse response functions at $q=500$ MeV, obtained from inversion of the corresponding Euclidean response are shown in
Fig.~\ref{fig:4he_xx}. The inversions are, to a very large degree, insensitive to the choice of default model response~\cite{Lovato:2015} needed by the maximum entropy method. Results obtained with one-body only (dashed line) and (one+two)-body (solid line) currents are compared with an analysis of the experimental world data~\cite{Carlson:2002} (empty circles).  The band provides an estimate for the dependence of the full results on the adopted default model, which appears to be remarkably small.
There is excellent agreement between the full theory and experiment. Two-body currents play a minor role in the longitudinal response function. On the other hand, they significantly enhance the transverse response function, not only in the dip region, but also in the quasi-elastic peak and threshold regions, providing the missing strength that is needed to reproduce the experimental results. It is worth noting that the width of the quasi-elastic peak seems to be correctly reproduced-- the nonrelativistic Fermi gas fails to predict this quantity at momentum transfers $q\simeq 500$ MeV. Thus, even at these relatively high momentum and energy transfers, the nonrelativistic dynamical framework may be more robust than comparisons between nonrelativistic and relativistic Fermi gas models would lead one to conclude \cite{DePace:2003}.

Finally, a direct evaluation of the $^{12}$C response functions via these same methods and with the same accuracy as the one of $^4$He shown in Fig. 3 has recently been obtained \cite{Lovato:2016} for $q=570$ MeV by using about 60 million core hours of computing time on Mira, the leadership class computing facility hosted at Argonne National Laboratory. Preliminary results show an enhancement of the transverse electromagnetic response function consistent with the $^4$He case. On the basis of the present $^4$He and $^{12}$C calculations, a consistent picture of the electroweak response of nuclei emerges, in which two-body terms in the nuclear electroweak current are seen to produce significant excess transverse strength from threshold to the dip region and beyond. Such a picture is at variance with the conventional one of inclusive quasi-elastic scattering, in which single-nucleon knockout is expected to be the dominant process in this regime.

\section*{Acknowledgement}
This work is based on results obtained in collaboration with S. Gandolfi, J. Carlson, S.C. Pieper, and R. Schiavilla. I am grateful to O. Benhar and N. Rocco for a number of discussions related to the subject of this paper.
This work was supported by the U.S. Department of Energy, Office of Science, Office of Nuclear Physics, under contract DE-AC02- 06CH11357. Under an award of computer time provided by the INCITE program, this research used resources of the Argonne Leadership Computing Facility at Argonne National Laboratory, which is supported by the Office of Science of the U.S. Department of Energy under contract DE-AC02-06CH11357.

%
% BibTeX or Biber users please use (the style is already called in the class, ensure that the "woc.bst" style is in your local directory)
\bibliographystyle{jpsj}
 \bibliography{biblio.bib}
%
% Non-BibTeX users please use
%
%\begin{thebibliography}{}
%
% and use \bibitem to create references.
%
%\bibitem{RefJ}
% Format for Journal Reference
%Journal Author, Journal \textbf{Volume}, page numbers (year)
% Format for books
%\bibitem{RefB}
%Book Author, \textit{Book title} (Publisher, place, year) page numbers
% etc
%\end{thebibliography}

\end{document}